\documentclass{aa}

\usepackage{natbib}
\usepackage{graphicx}
\usepackage{amsmath}
\usepackage{ulem}
\usepackage{amssymb}
\usepackage[colorlinks=true,linkcolor=blue,citecolor=blue]{hyperref}
\usepackage{txfonts}
\usepackage{xparse,xcolor}

\begin{document}

\title{Modelling rebrightenings, reflares, and echoes in dwarf nova outbursts}

\author{J.-M. Hameury\inst{1}
    \and
          J.-P. Lasota\inst{2,3}
}

\institute{
     Observatoire Astronomique de Strasbourg, Université de Strasbourg, CNRS UMR 7550, 67000 Strasbourg, France \\
                    \email{jean-marie.hameury@astro.unistra.fr}
\and
     Institut d'Astrophysique de Paris, CNRS et Sorbonne Universit\'e, UMR 7095, 98bis Bd Arago, 75014 Paris, France
\and
     Nicolaus Copernicus Astronomical Center, Polish Academy of Sciences, Bartycka 18, 00-716 Warsaw, Poland
}

   \date{Received / Accepted}


  \abstract
  {The disc instability model (DIM) accounts well for most of the observed properties of dwarf novae and soft X-ray transients, but the rebrightenings, reflares, and echoes  occurring at the end of outbursts or shortly after  in WZ Sge  stars or soft X-ray transients have not yet been convincingly explained by any model.}
   {We determine the additional ingredients that must be added to the DIM to account for the observed rebrightenings.}
   {We analyse in detail a recently discovered system, TCP~J21040470+4631129, which has shown very peculiar rebrightenings, model its light curve using our numerical code including mass transfer variations from the secondary, inner--disc truncation, disc irradiation by a  hot white dwarf and, in some cases, the mass-transfer stream over(under)flow.}
   {We show that the luminosity in quiescence is dominated by a hot white dwarf that cools down on time scales of months. The mass transfer rate from the secondary has to increase by several orders of magnitudes during the initial superoutburst for a reason that remains elusive, slowly returning to its secular average, causing the observed succession of outbursts with increasing quiescence durations, until the disc can be steady, cold, and neutral; its inner parts being truncated either by the white dwarf magnetic field or by evaporation. The  very short, quiescence phases  between reflares are reproduced when the mass-transfer stream overflows the disc. Using similar additions to the DIM, we have also produced light curves close to those observed in two WZ Sge stars, the prototype and EG Cnc.}
   {Our model successfully explains the reflares observed in WZ Sge systems. It requires, however, the inner disc truncation in dwarf novae to be due  not (only) to the white dwarf magnetic field but, as in X-ray binaries, rather to evaporation of the inner disc. A similar model could also explain reflares observed in soft X-ray transients.}

   \maketitle
%

\section{Introduction}
Dwarf novae (DNe) are cataclysmic variables undergoing regular outbursts \cite[see][for a full review of these systems]{w03}, that are well explained by the disc instability model (DIM). In this model, a thermal--viscous instability is triggered  in the accretion disc when its temperature enters the range in which hydrogen is partially ionized and opacities are strongly temperature-dependent \citep[see][for reviews of the model]{l01,h20}. The DIM successfully explains many of the observed properties of DNe: their outburst amplitudes, durations, and recurrence times. It also accounts well for the observed dichotomy between stable and unstable systems: the critical mass transfer rate separating dwarf novae from nova-like variables as deduced from observations corresponds precisely to the predictions of the model \citep{dol18}. Its major drawback is the description of angular momentum transport that essentially relies on the \citet{ss73} prescription, which parametrizes our poor understanding of the exact mechanisms responsible for angular momentum transport. The magnetorotational instability \citep{bh91} is certainly one of them, but other mechanisms must be at work, at least in quiescence \citep[see, e.g.][]{sldf18}.

\begin{figure*}
\includegraphics[width=\textwidth]{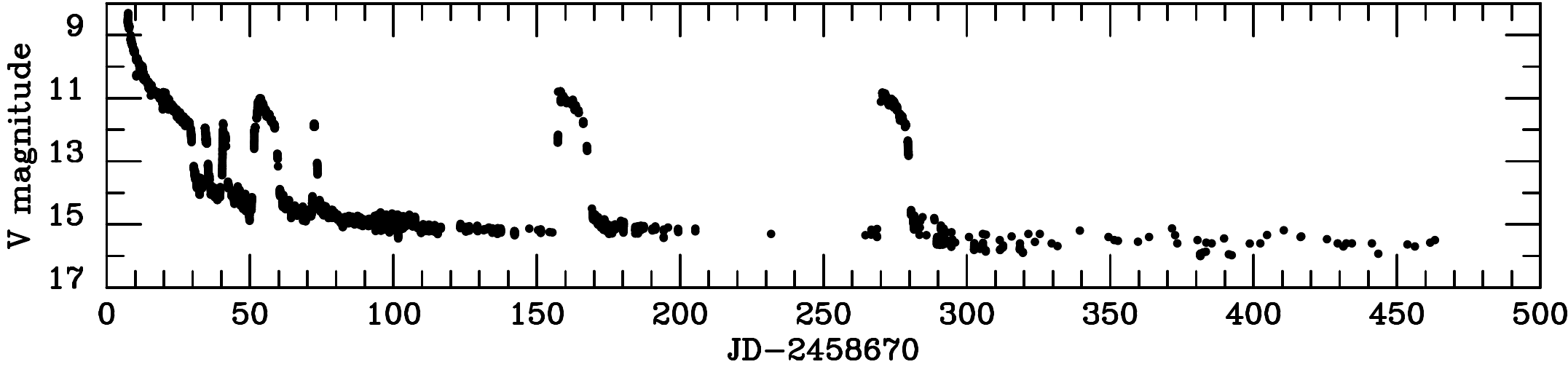}
\caption{Light curve of TCP~J2104 (data from AAVSO).}
\label{fig:lc}
\end{figure*}

Several additional ingredients had, however, to be incorporated in the model to account for properties of various subclasses of DNe. For example, variations of the mass transfer rate are required to explain the Z Cam systems that alternate between bright, steady periods, and periods during which they undergo DN outbursts. Truncation of the inner disc by a magnetic field is needed to model the DN phenomenon in intermediate polars \citep{hl17}, but truncation (not necessarily magnetic) is also needed to explain outbursts in other sources, such as  SS~Cyg \citep{shl03} and WZ Sge stars that we discuss further in this section, or soft X-ray transients (SXTs) that are  low-mass X-ray binary analogues of DNe. Irradiation of the outer accretion disc by radiation emitted by the inner portions of the disc is an essential ingredient for modelling SXTs \citep{vp96}, but irradiation may also be important in cataclysmic variables \citep{hld99}; in the latter case, it is mostly the inner-disc irradiation by the hot (accretion heated) white dwarf that is important..

The SU UMa stars are DNe that show, in addition to normal outbursts, longer and brighter outbursts, called superoutbursts that occur regularly, typically every few normal outbursts. The tidal-thermal instability model, proposed by \citet{o89} invokes the existence of a tidal instability that is supposed to occur when the mass ratio is low enough for the 3:1 resonance radius to fit inside the Roche lobe of the primary. When the disc outer radius reaches this resonance radius, the disc becomes eccentric and precesses \citep{w88}, generating periodic fluctuations of the luminosity, the so-called superhumps, at a period that is slightly larger than the orbital period. \citet{o89} proposed that the tidal instability also generates higher viscous dissipation in the whole disc which lasts until the disc has shrunk enough and becomes circular again. This TTI model has been challenged by the observation in 1985 of a superoutburst and superhumps in U Gem \citep{sw04}, which is a normal DN system that cannot be subject to the tidal instability. This led \citet{s05} to propose that the U Gem superoutburst was caused by an enhancement of the mass transfer rate by a factor 20--50, and that superhumps in U Gem are caused by periodic irradiation of the secondary \citep{s09}. Indeed, \citet{hlw00} showed that the combination of disc irradiation, mass transfer variations and disc truncation can naturally account for the observed variety of DN light curves, including the alternation of long and short outbursts of the SU UMa stars. It is also worth adding that the physics of the viscosity--increase mechanism in the TTI remains a total mystery.

The WZ Sge systems are SU UMa stars that exhibit only superoutbursts \citep[see][for a review of this subclass]{k15}. Their outbursts are very rare, with a recurrence time of the order of decades (30 years for example in the case of WZ Sge, the prototype of this subclass) and they are often followed by rebrightenings that occur shortly after, or even at the end of a superoutburst. Until now, no model has been able to convincingly explain reflares, which are similar to those observed in some X-ray transients, for which also no credible explanation has ever been given. Several suggestions have been made in the past to explain rebrightenings, a detailed list of which can be found in \citet{k15}. These include a temporary increase of the mass transfer rate as a consequence of irradiation of the secondary \citep{aks93,hlw00}; an exponential decay of the viscosity in the disc between rebrightening accompanied by a partial revival of the viscous processes during these rebrightenings \citep{omm01}; the existence of a mass reservoir beyond the 3:1 resonance radius \citep{uak08}; and the decoupling of the tidal and thermal instability that might occur in systems with a low mass ratio \citep{h01}. More recently, \citet{mm15} proposed that rebrightenings are due to the back and forth propagation of heating and cooling waves; their model, however, requires the viscosity in the cold state to be roughly the same as in the hot state, and, because the inner disc is always hot, there is no real quiescence period.

The extraordinary light curve of TCP~J21040470+4631129, hereafter TCP~J2104, sheds a new light on these systems. This source was unnoticed until 2019, when it underwent a large outburst, during which it reached $V=8.5$, and then had six outbursts in less than one year, including three superoutbursts; its light curve in shown in Fig. \ref{fig:lc}, which extends the light curve provided by \citet{tni20} by nine months and includes the last observed superoutburst. No outburst has been recorded since this one which occurred in April 2020.

In this paper, we propose model which consistently describes the three main phases of the  TCP~J2104 multiple outburst: the first, major superoutburst, the following outbursts (rebrightenings) and quiescence. We suggest that the first, major outburst of TCP~J2104 is caused by a sharp enhancement of the mass transfer, which, maintained by irradiation of the secondary star, results in a superoutburst. When the superoutburst ends, the mass transfer rate $\dot{M}_{\rm tr}$ is still large because the white dwarf has been heated to high temperatures; $\dot{M}_{\rm tr}$ is close to the critical value for the disc to be unstable, and a sequence of closely spaced outbursts occurs with an alternation of long and short outbursts. As the mass transfer rate slowly decreases, presumably because of the cooling of the white dwarf, the time interval between outbursts increases; if the accretion disc is truncated, for example, by a magnetic field, it finally becomes stable when the mass transfer rate is low enough for the disc to be neutral everywhere, as has been proposed by \citet{hlh97} for WZ Sge. We then extrapolate our model to other transients that show rebrightenings, the WZ Sge systems and the soft X-ray transients. We show that the model applies to the WZ Sge stars, as we are able to reproduce light curves similar to those of WZ Sge itself and EG Cnc. 

As a note of caution, we emphasize that our aim here is not to reproduce the very details of light curves of the systems we intend to explain, because the physical processes involved in our model are taken into account in a crude way and we want to minimize the number of free parameters that are not constrained by observations.  

For this reason, although we argue that reflares in SXTs could also be well reproduced by our model, we refrain from doing so because outer-disc X-ray irradiation, crucial in SXTs, adds one or more additional free parameters that are not well constrained \citep{tdl18}. Reproducing SXTs light curves would then bring little additional insight on the physics involved in the outbursts of these systems.

\section{Observations of TCP~J21040470+4631129}

\begin{table}
\caption{Outburst characteristics}
\begin{tabular}{lccc}
\hline \hline
Outburst & $t_{\rm q}$ & duration & $V_{\rm peak}$ \\
\hline
1st SO & & $\sim 25$ & 8.5 \\
1st NO & 3.5 & 1.6 & 11.9 \\
2nd NO & 5.0 & 1.6 & 11.8 \\
2nd SO & 9.2 & 8.5 & 11.0 \\
3rd NO & 12.2 & 1.4 & 11.8 \\
3rd SO & 83.2 & 10.9 & 10.8 \\
4th SO & 101.8 & 10.5&  10.8 \\
\hline
\end{tabular}
\end{table}

This system is located at a distance $d=109 \pm 1.4$~pc determined by {\it GAIA} \citep{g18}. Its orbital period is $P_{\rm orb} = 1.28$~hr, that is 77 minutes, close to the minimum period of cataclysmic variables \citep{nbb19}. The mass ratio $q=M_2/M_1$, where $M_1$ and $M_2$ are the primary and secondary masses respectively, measured in solar units, is estimated by \citet{tni20} to be $q=0.088$; this estimate is based on the method proposed by \citet{ko13} who use the relation between $q$ and the period of superhumps in their early stage. Other determinations for q range from $q=0.1$ \citep{nbb19} to $q=0.062$ \citep{tni20} depending on the particular relation between the superhump period and $q$ that one uses and on the stage at which superhumps are considered. There is thus a significant uncertainty on $q$. In the following, we use $M_2=0.1$, and we consider two possible values for $M_1$: 0.6 and 1.

Figure \ref{fig:lc} shows the light curve in the $V$ band of TCP~J2104, using the AAVSO data. We consider only observations that provide data either in the $V$ or $CV$ band, for which the error on magnitude is smaller than 0.05. We average magnitudes when observations are separated by less than 0.005~d, that is 7 min. During the overlapping time interval this light curve is identical to that provided by \citet{tni20}. 

The outburst characteristics (quiescence time between two outbursts, $t_{\rm q}$, outburst duration, peak magnitude) are given in Table 1. The duration of an outburst is defined as the time during which the system is brighter than $V=13$, and the quiescence is defined by $V > 13$. The distribution of durations is clearly bimodal: four outbursts, including the initial outburst, last for more than eight days, whereas the remaining three last for less than two days. In the following, the long outbursts are labelled as ``superoutbursts'' and short ones are designated as ``normal outbursts''. Superhumps were also detected during these long outbursts, making them bona fide SU UMa type superoutbursts. \citet{tni20} note that late superhumps were detected until the third normal outburst which, they suggest, indicates that the disc remained eccentric for about forty days after the end of the initial outburst \citep[see, however,][]{s09,s13}. In addition to the six major rebrightenings mentioned in this table, a faint, stunted outburst with a 0.6 mag amplitude occurred between the second normal outburst and the second superoutburst.

\begin{figure}
\includegraphics[width=\columnwidth]{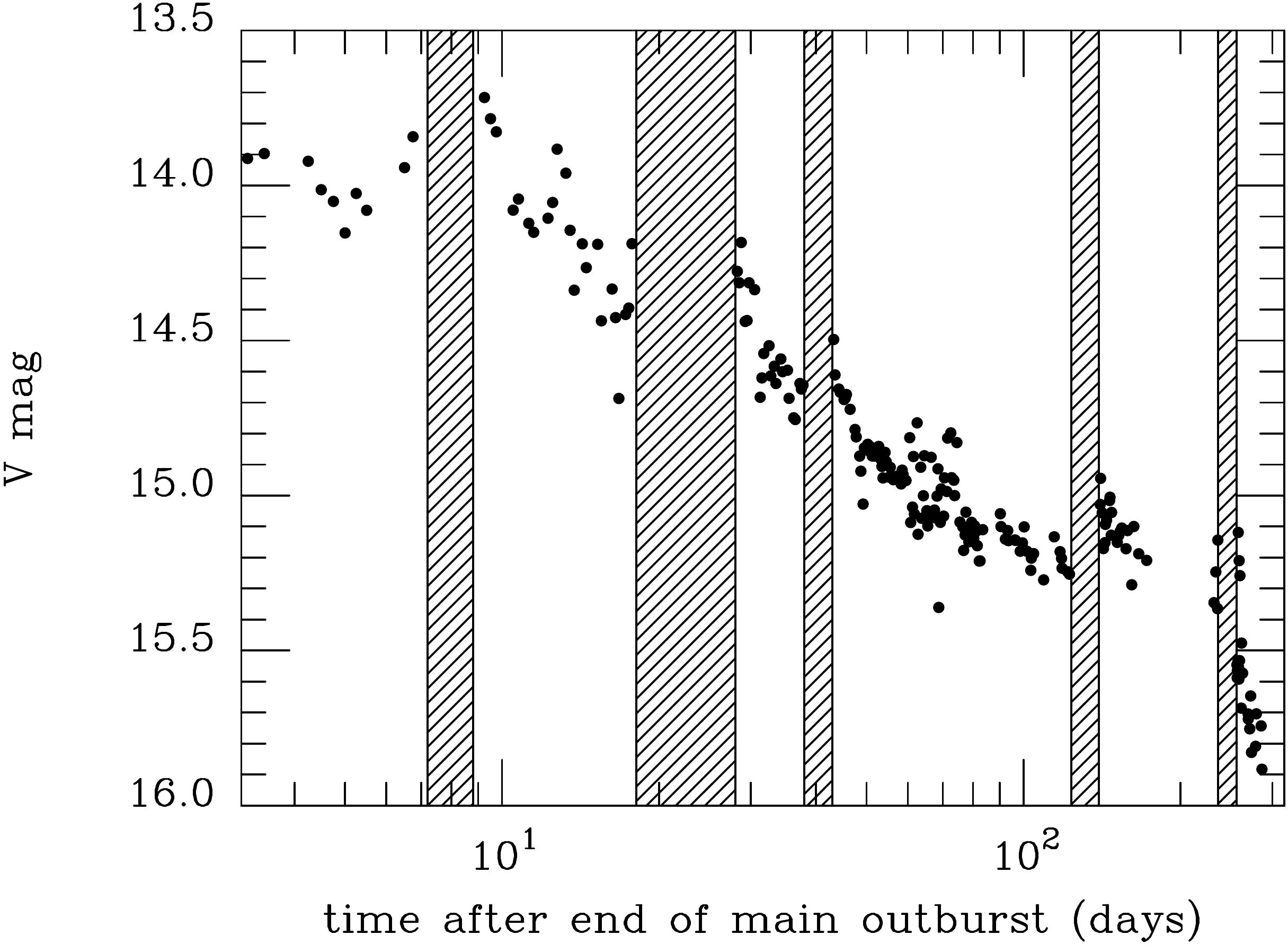}
\caption{Variation of the quiescent luminosity of TCP~J2104. The hatched areas correspond to outbursts. (Data from AAVSO)}
\label{fig:quiescence}
\end{figure}

Figure \ref{fig:quiescence} shows that the magnitude during quiescence slowly decays with time, being about $V=14$ after the first superoutburst and decaying to $V=15.5 - 16$ at the end of the observation period; this is still significantly brighter than the pre-outburst luminosity. It is difficult to derive from the observations an accurate time--dependence of the quiescent optical flux; Fig. \ref{fig:quiescence} indicates that a power law with index 0.4, possibly with a cut-off on a time-scale of the order of a year, would be a good fit to the data. Over the 4.5~yr interval before the main outburst during which TCP~J2104 was observed by {\it GAIA}, it brightened by 0.5 mag from 17.92 to 17.46 \citep{tni20}. During this time interval, no outburst was detected, neither by {\it GAIA} nor by the All Sky Automated Survey for SuperNovae (ASAS-SN) \citep{tni20}, despite a dense coverage of this interval indicating that it is extremely unlikely that an outburst occurred during these 4.5~yr.

Based on the red colour ($G_{\rm BP} - G_{\rm RP} = 0.58$), \citet{tni20} assume that, during quiescence, the optical light originates from the secondary. The Roche-lobe filling secondary's radius is $7.5 \times 10^9$~cm for a mass of 0.1~M$_\odot$, weakly dependent on $M_2$. Even if its temperature were as high as 3000~K, its magnitude would be $V=19.4$. In order to have $V=18$, one needs a secondary's temperature of 3500~K. The secondary cannot therefore contribute during quiescence. On the other hand, the contribution of the white dwarf is significant; a 0.6 M$_\odot$ white dwarf at a distance of 109~pc and with a temperature of 8000~K has $V=18.1$, assuming a blackbody spectrum. If the mass is 1.0~M$_\odot$, the same magnitude is obtained for $T=11,000$~K. This sets an upper limit on the white dwarf temperature, because, as we shall see later, the contribution of the irradiated disc is significant and explains in part the red colour of this system. The white-dwarf temperatures we deduce are fully compatible with those observed in CVs at periods close to the minimum \citep{sion12}.

\begin{figure}
\includegraphics[width=\columnwidth]{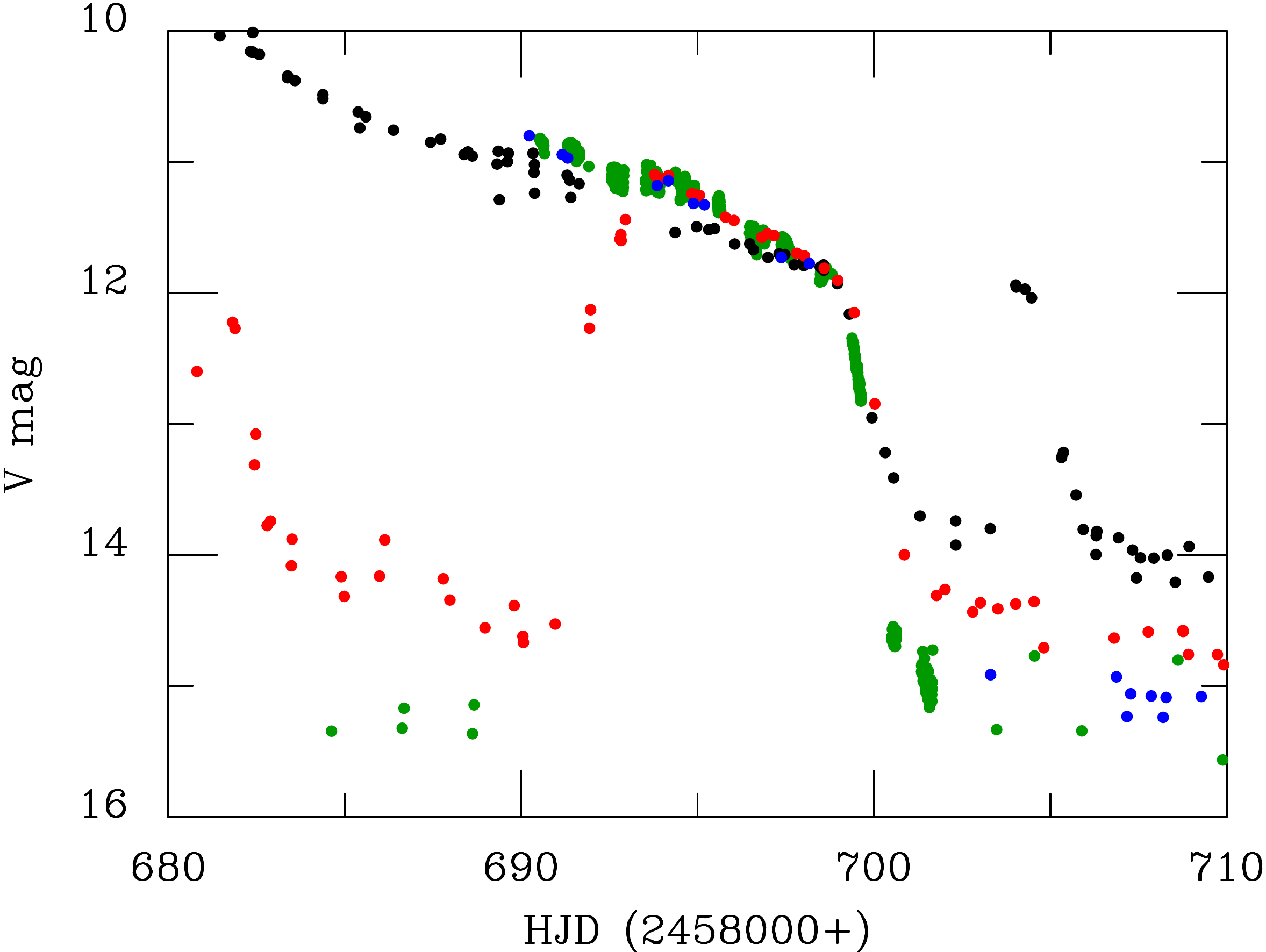}
\caption{Initial outburst superimposed on the 3 following superoutbursts. The time axis refers to the initial outbursts, the subsequent outbursts have been shifted so that their breaks in the light-curve shape (start of the cooling front) coincide. Black, red, blue, and green dots correspond to the initial, first, second, and third superoutburst respectively.}
\label{fig:su}
\end{figure}

\begin{figure}
\includegraphics[width=\columnwidth]{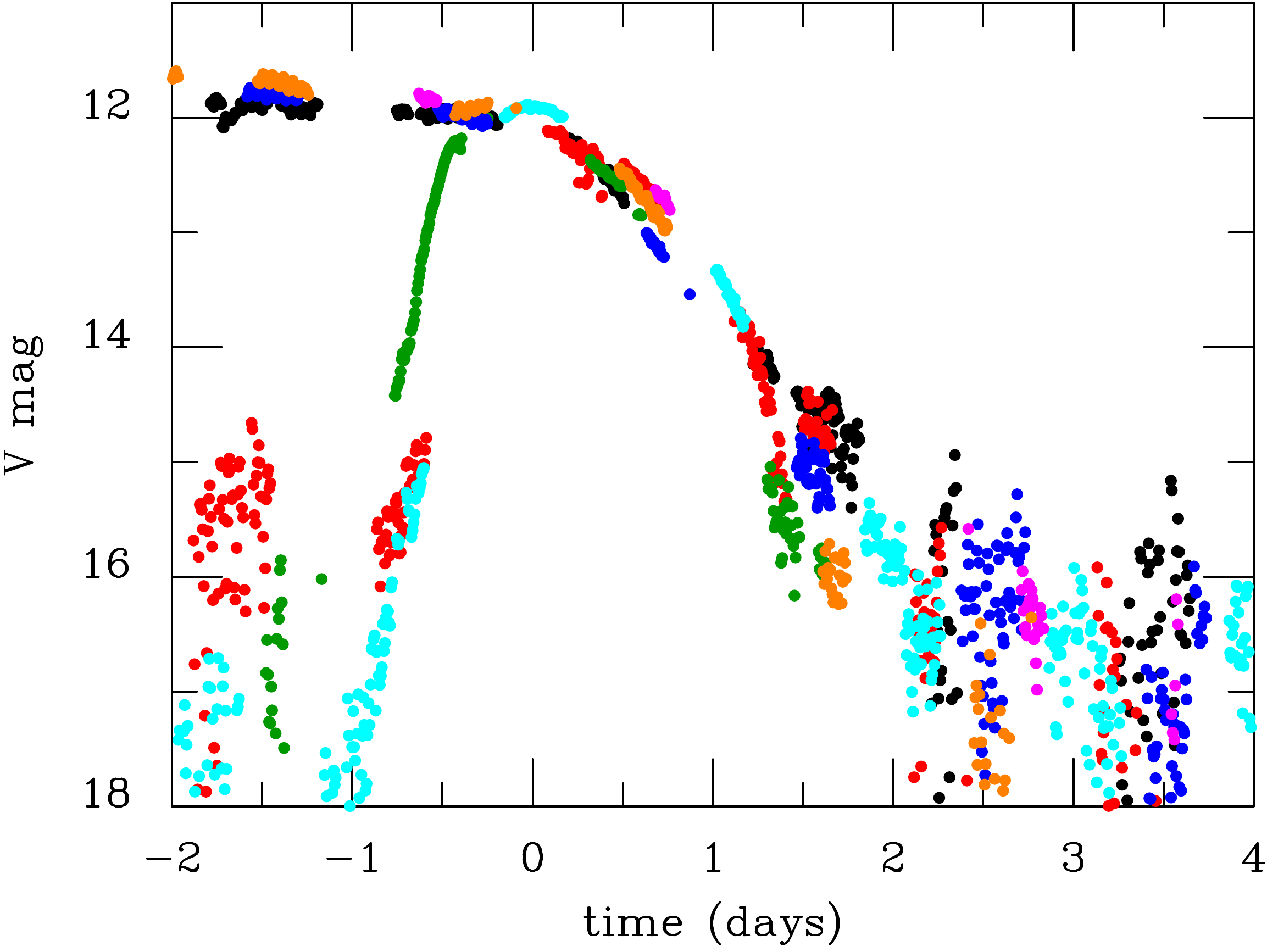}
\caption{Steep outburst decline corresponding to the propagation of cooling fronts for all observed outbursts. A rough estimate of the contribution of the post-outburst luminosity has been subtracted.}
\label{fig:fronts}
\end{figure}

The light curves of superoutbursts show two phases of decline from maximum. The first one is slow and corresponds to viscous decay at a rate depending solely on the outer disc size. The second one is faster, begins after a break in slope and is produced by a cooling wave propagating inwards from the outer disc edge. This wave brings the the whole disc to a cold state: the quiescence.
Figure \ref{fig:su} shows a detailed view of the initial superoutburst superimposed on the three following superoutbursts in such a way that the rapid decline phase of each outburst starts at the same time. The properties of the superoutbursts are remarkably similar. The slopes of the slow decline phase during the four superoutbursts are almost identical; the magnitude at the break that occurs when a cooling front is initiated is also the same for all four superoutbursts, indicating that the disc extension is the same. During the slow decay phase, the disc is, as we shall see later, fully on the hot branch, and close to being steady. The superoutburst durations are different, though, as well as the peak luminosities. The post-outburst magnitude increases from outburst to outburst, as was already clearly visible in Figs. \ref{fig:lc} and \ref{fig:quiescence}; as mentioned earlier, the most obvious explanation for this is the cooling of the white dwarf.

The sharp decline phase is also very similar for all outbursts, including normal outbursts. This phase corresponds to the propagation of a cooling front starting from the outer disc edge throughout the disc. Figure \ref{fig:fronts} shows this phase for all outbursts once an estimate of the quiescent luminosity, due to the white dwarf with a possible contribution from reprocessing of its light by the disc has been removed. This indicates that the velocity of the cooling front is the same for a given magnitude, or, equivalently for a given position of the front. Since the front velocity depends mainly on the viscosity in the hot state \citep{mhs99}, it means that this viscosity is identical at the end of all outbursts, whether normal or long.

The peak magnitude of normal outbursts is about the same as the magnitude at which a cooling front is initiated for superoutbursts, indicating that the disc size is also the same, or does not differ much, for normal outbursts and at the end of the decay phase of the superoutbursts. 

\section{Model}

In the following, we use the code developed by \citet{hmd98} and subsequently modified to include disc truncation by either a magnetic field or under the effect of evaporation, disc irradiation and tidal heating \citep{dhl01,bhl01}. As in \citet{hlw00}, we assume that irradiation of the secondary star by the accretion luminosity results in an increase of the mass transfer rate $\dot M_{\rm tr}$. For simplicity, we set $\dot M_{\rm tr}$ proportional to the accretion rate:
\begin{equation}
\dot M_{\rm tr} = \max (\gamma \dot M_{\rm acc},\dot M_0),
\label{eq:mdot_tr}
\end{equation}
where $\dot M_0$ is the mass-transfer rate in absence of irradiation and
 $\gamma < 1$ for obvious stability reasons.

We calculate the visual magnitude of the disc assuming local blackbody spectra, and we use the formula (6) from \citet{s89} to account for the angular dependence of the disc emissivity that also includes limb darkening:
\begin{equation}
L(i) = L_{0} \cos i \, (0.4 + 0.6 \cos i),
\end{equation}
where $L(i)$ and $L_{0}$ are the luminosities at angles $i$ and 0 respectively; this adds an extra term $0.4 + 0.6 \cos i$ to the standard $\cos i$ term, which becomes important at high inclinations.

As discussed above, the outburst cycle of TCP~J2104 can be decomposed into three phases: the first long superoutburst, the subsequent outbursts and the quiescent epochs. Since different physical phenomena dominate or characterize each phase, we first discuss separately the models of each phase and in Sect. \ref{sec:fullc} we then present the synthesis modelling the full outburst light curve.

\subsection{Initial superoutburst}
\label{sec:initial}

\begin{figure}
\includegraphics[width=\columnwidth]{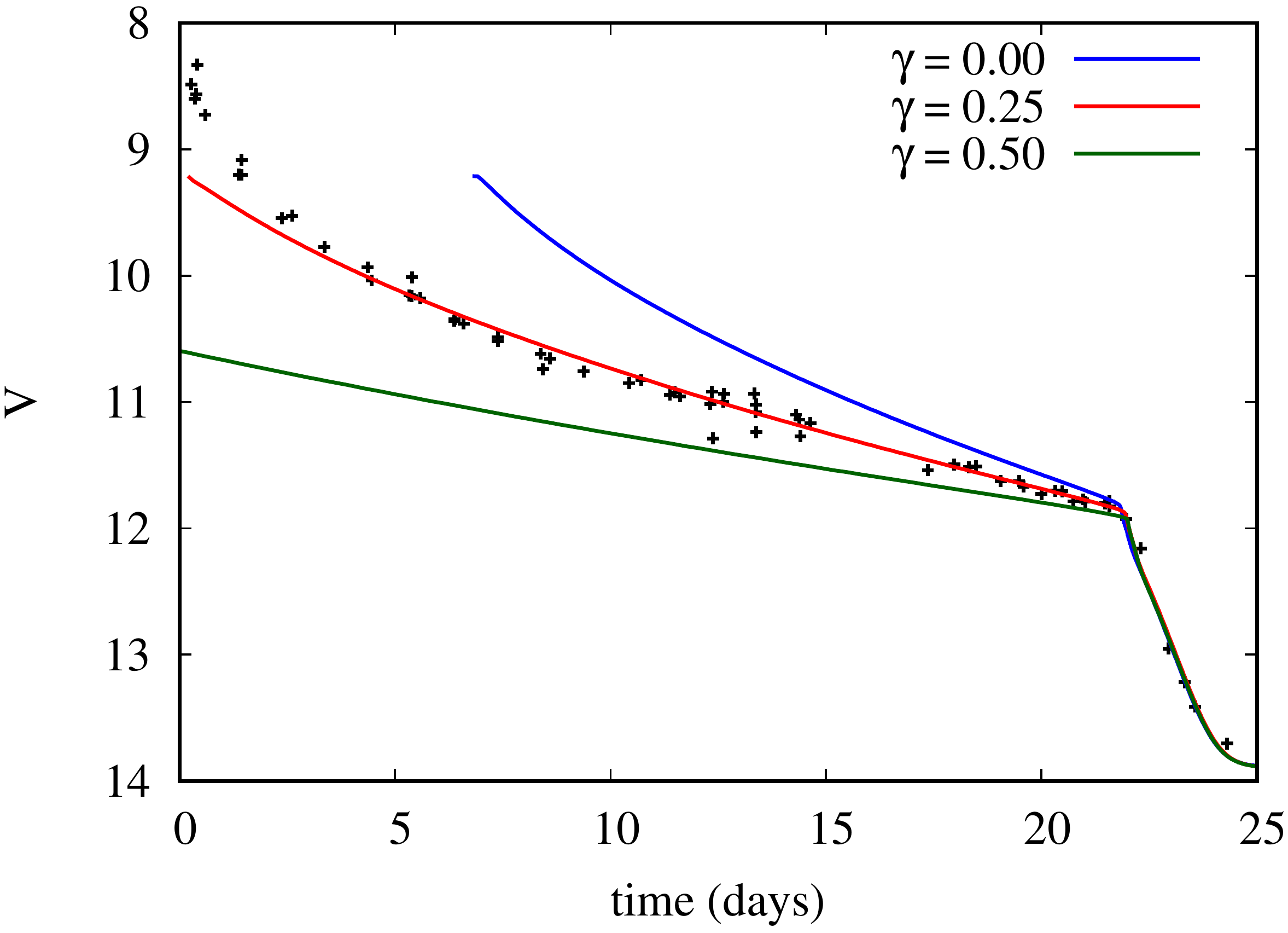}
\caption{Observed light curve in $V$ (points) compared to models assuming $\alpha_{\rm h}=0.2$ and $\gamma=0$ (blue), 0.25 (red) and 0.5 (green).}
\label{fig:su0}
\end{figure}

\begin{figure}
\includegraphics[width=\columnwidth]{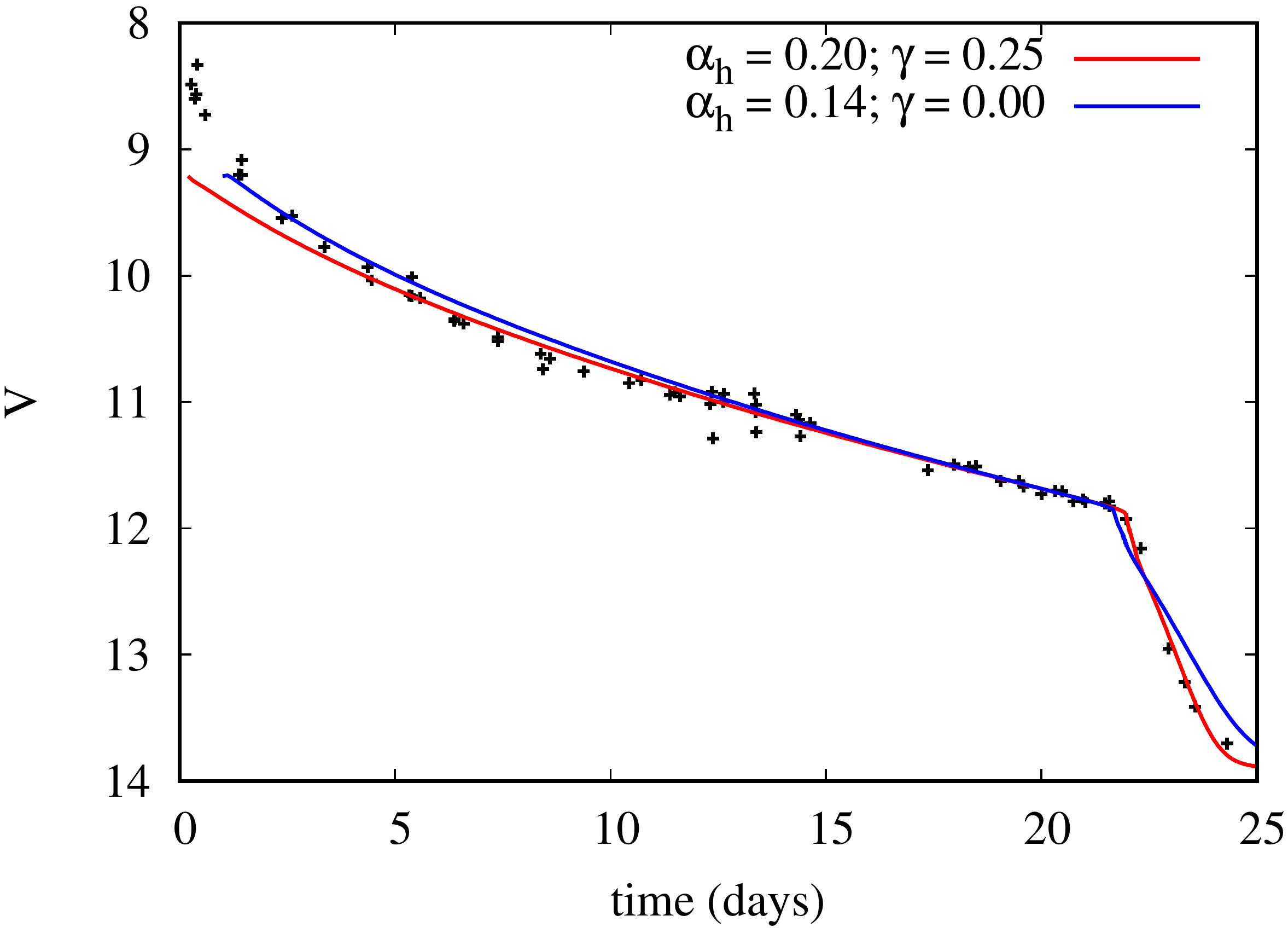}
\caption{Same as Fig. \ref{fig:su0}, with the model shown in Fig. \ref{fig:su0} fitting the data and a model in which $\gamma=0$ and $\alpha_{\rm h}=0.14$ which fits equally well the outburst maximum.}
\label{fig:su0a}
\end{figure}

During the long, initial outburst, the disc is at any time not far from being steady, so that its subsequent evolution is almost independent of the initial conditions. We therefore choose to follow the evolution of an initially steady disc, with a high mass transfer rate which is turned off at time $t=0$, except for the irradiation-driven term, given by Eq. (\ref{eq:mdot_tr}).

At this point, the cause of the outburst is not specified. As shall be discussed in Sect. \ref{sec:quiesc} and \ref{sec:fullc}, it is most probably a a strong mass--loss enhancement intrinsic to the secondary star, possibly strengthened by irradiation of the secondary star once the instability has started. The subsequent evolution of the system is independent of the initial trigger, precisely because the disc is not far from being steady. It does, however, depend on how the mass transfer rate returns back to its secular mean, which includes the effect of the secondary irradiation.

Figure \ref{fig:su0} shows the time evolution of a system with the same orbital parameters as TCP~J2104: $P_{\rm orb}=1.28$~h, $M_1=0.6$, $M_2=0.08$, $d=109$~pc. The viscosity parameter that must be different in outburst and in quiescence \citep{hmd98} was taken to be respectively  $\alpha_{\rm h} = 0.2$  and $\alpha_{\rm c} = 0.04$. 

The slow decline phase is well approximated by a model in which $\gamma = 0.25$ and $\alpha_{\rm h} = 0.2$. The rapid decline occurring when a cooling front propagates is also well reproduced by the model. It is also possible to reproduce the slow decline if $\gamma = 0$; one then requires $\alpha_{\rm h}= 0.14$. This model does not reproduce as well the rapid decline: the slope is shallower than observed, even though the disagreement is not strong. It is therefore possible to reproduce the light curve of the initial outburst without a mass transfer enhancement following the initial mass transfer burst; as we shall see in the next section, such a mass transfer enhancement is nevertheless requested to account for the rebrightenings.

If the hot viscosity parameter in TCP~J2104 is close to 0.2  \citep[according to][$\alpha_h$ is between 0.1 and 0.2 in CVs ]{kl12}, the actual mass transfer rate during the slow-decay phase must be comparable to the mass accretion rate, so that the disc evolves on a time scale that is slightly longer than the viscous time. At this phase the contribution of the white dwarf is not taken into account; as discussed later, it becomes significant when the disc magnitude is 13 -- 14 or fainter; the last data point in Figs. \ref{fig:su0} and \ref{fig:su0a} therefore contains a significant contribution from the white dwarf. 
 
The break in the light curve at $V \sim 12$ occurs when the mass accretion rate becomes smaller than the minimal value allowed on the hot, stable branch, $\dot{M}_{\rm crit}^+$ determined at the outer edge of the disc, and a cooling front is initiated. As $\dot{M}_{\rm crit}^+$ depends only on the primary mass and the disc outer radius, but not on the viscosity parameters, the magnitude at the break is rather well defined. The outer disc radius is given by the tidal truncation radius, $r_{\rm tid}$; we take $r_{\rm tid} = 0.9 R_{\rm L1}$, where $R_{\rm L1}$ is the spherical Roche radius of the primary, for which we use the classical \citet{e83} formula. For $M_1=0.6$, one gets $r_{\rm tid} = 1.83 \times 10^{10}$~cm, and $r_{\rm tid} = 2.28 \times 10^{10}$~cm for $M_1=0.6$. We find that the break occurs at the observed $V=11.9$ for $M_1=0.6$, and at $V=11.4$ if $M_1=1.0$, with $\cos i=0.45$ in both cases. This actually sets the value of $\cos i$ if the primary mass is known. 

\subsection{Rebrightenings}
\label{sec:rbr}

If the outburst is of the inside-out type, the quiescence time is independent of the mass transfer rate and is given by \citep{l01}: 
\begin{equation}
t_{\rm q} = 130 \; \delta \left( \frac{\alpha_{\rm c}}{0.01}\right)^{-1} \left( \frac{T_{\rm c}}{3000 \; \rm K} \right)^{-1} \left( \frac{r_{\rm out}}{10^{10} \; \rm cm}\right)^{0.5} M_1^{0.5} \; \rm d,
\end{equation}
where $\delta$ is a logarithmic term of the order of unity, $T_{\rm c}$ is the mid-plane disc temperature, and $r_{\rm out}$ the outer disc radius. For the parameters of TCP~J2104, $t_{\rm q}$ is of the order of 35 days for $\alpha_{\rm c}=0.04$. This is much longer than the first quiescent intervals. 

In the case of outside--in outbursts, $t_{\rm q}$ is given by:
\begin{equation}
t_{\rm q} = 0.55 \left( \frac{\alpha_{\rm c}}{0.01}\right)^{-1} \left( \frac{T_{\rm c}}{3000 \; \rm K} \right)^{-1} \left( \frac{\dot M_{\rm tr}}{\dot M_{\rm crit}^+}\right)^{-2} \left( \frac{r_{\rm out}}{10^{10} \; \rm cm}\right)^{0.44} M_1^{0.52} \; \rm d .
\end{equation}
$\dot M_{\rm tr} / \dot M_{\rm crit}$ must be less than unity, but cannot be much smaller than one for the outburst to be of the outside--in type. Short quiescence times, of the order of a few days can easily be obtained and account for the observed short interval between reflares.

This requires the mass transfer rate during the rebrightening period to be close to the critical rate at the end of the first superoutburst, which is much higher than the secular mass transfer rate. This high mass transfer rate described by a simple formula such as Eq. (\ref{eq:mdot_tr}) because of the smallness of the quiescent accretion rate; some other mechanism must be at work to account for the persistence of a high mass transfer rate between reflares; it can be intrinsic to the secondary star and related to the mechanism that produced the initial mass transfer outburst, or it could be related to the heating of the secondary surface by the very hot white dwarf. In the latter case, one should, in principle, be able to establish a relation between the mass transfer rate and the white dwarf luminosity; it turns out, however, that establishing such a relation in a reliable way from first principles would involve so many unconstrained parameters and physical effects that it is doomed to failure \citep[see e.g.][]{sm04,vh07}.

\begin{figure}
\includegraphics[width=\columnwidth]{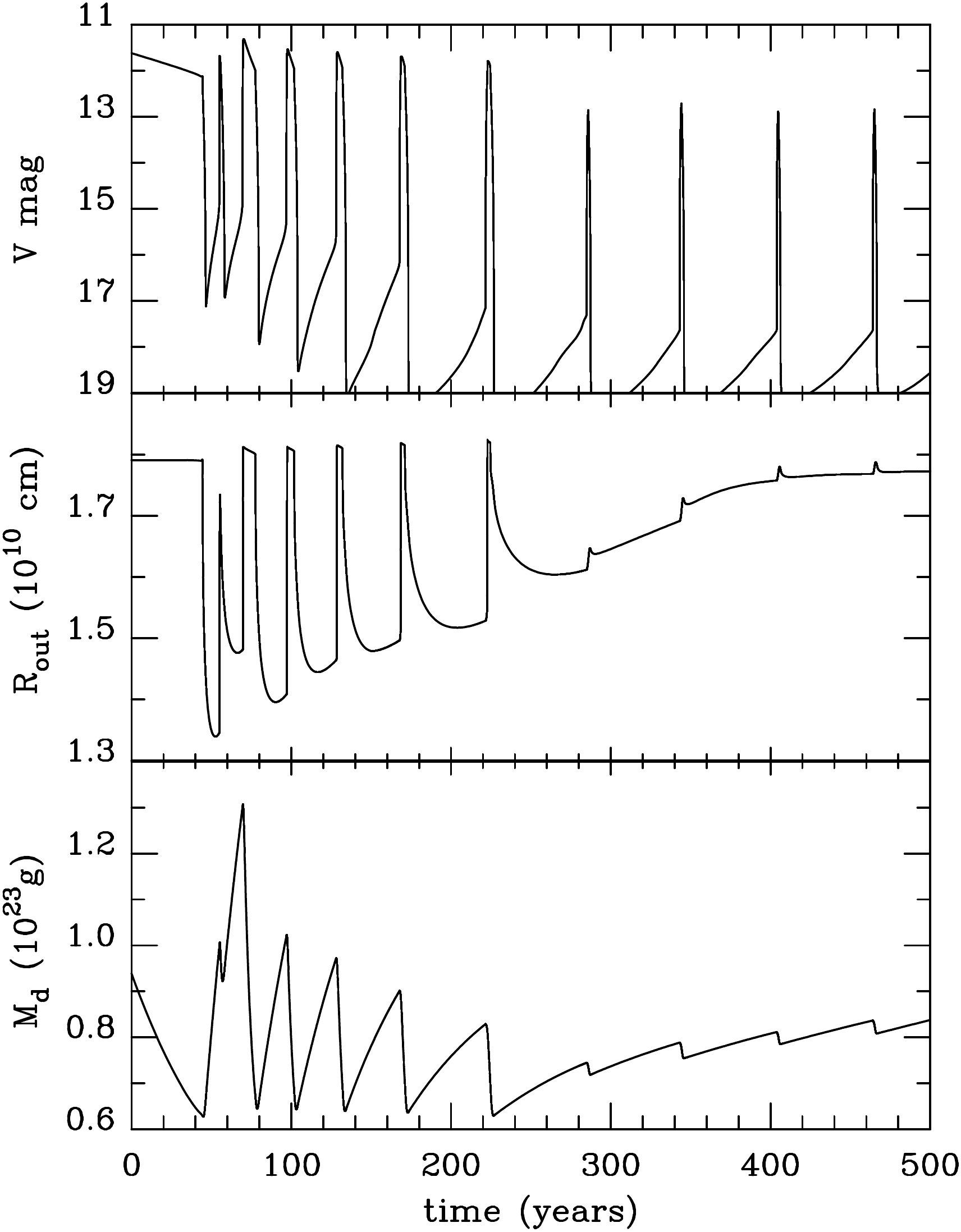}
\caption{Time evolution when the mass transfer rate exponentially decreases with an e-folding time of 58~d. {\it Top panel:} Visual magnitude of the disc only. {\it Intermediate panel:} outer disc radius. {\it Bottom panel:} disc mass}
\label{fig:model52}
\end{figure}

First, we consider the case when the initially enhanced mass-transfer rate decays exponentially after the end of the major superoutburst, with no additional contribution due to the irradiation of the secondary. Figure \ref{fig:model52} shows the evolution of a system in which the mass transfer rate decreases with time as $\dot{M}_{\rm tr} = 1.2 \times 10^{17} \exp [-t/(5 \times 10^6 \; \rm s)] + 10^{15}$~g~s$^{-1}$. The decay time scale of $5 \times 10^6$~s, that is 58~d,  used here is rather arbitrary, and was chosen because it corresponds approximately to the duration of the initial reflare sequence, up to the third normal outburst. We took $M_1=0.6$, $M_2=0.08$, $P_{\rm orb} = 1.28$, $\gamma = 0$, $T_{\rm wd}=0$, $\alpha_{\rm c}=0.04$, and $\alpha_{\rm h}=0.2$. After the main superoutburst, we obtain an initial sequence consisting of a normal outburst followed by five superoutbursts. This sequence is followed by weaker, normal outbursts as the mass transfer rate stabilizes at $10^{15}$~g~s$^{-1}$. The first four outbursts are outside-in; the instability is triggered at a distance decreasing from $7 \times 10^9$~cm from the white dwarf for the first outburst to $1.5 \times 10^9$~cm for the fourth outburst. During the initial six outbursts following the main superoutburst, the heating front reaches the outer edge of the disc which expands to the tidal truncation radius; once a cooling front has started from the outer edge, the disc contracts rapidly. The heating front does not, by far, reach the outer disc edge during the last four outbursts visible in Fig. \ref{fig:model52}, so that the disc radius remains almost constant.

\begin{figure}
\includegraphics[width=\columnwidth]{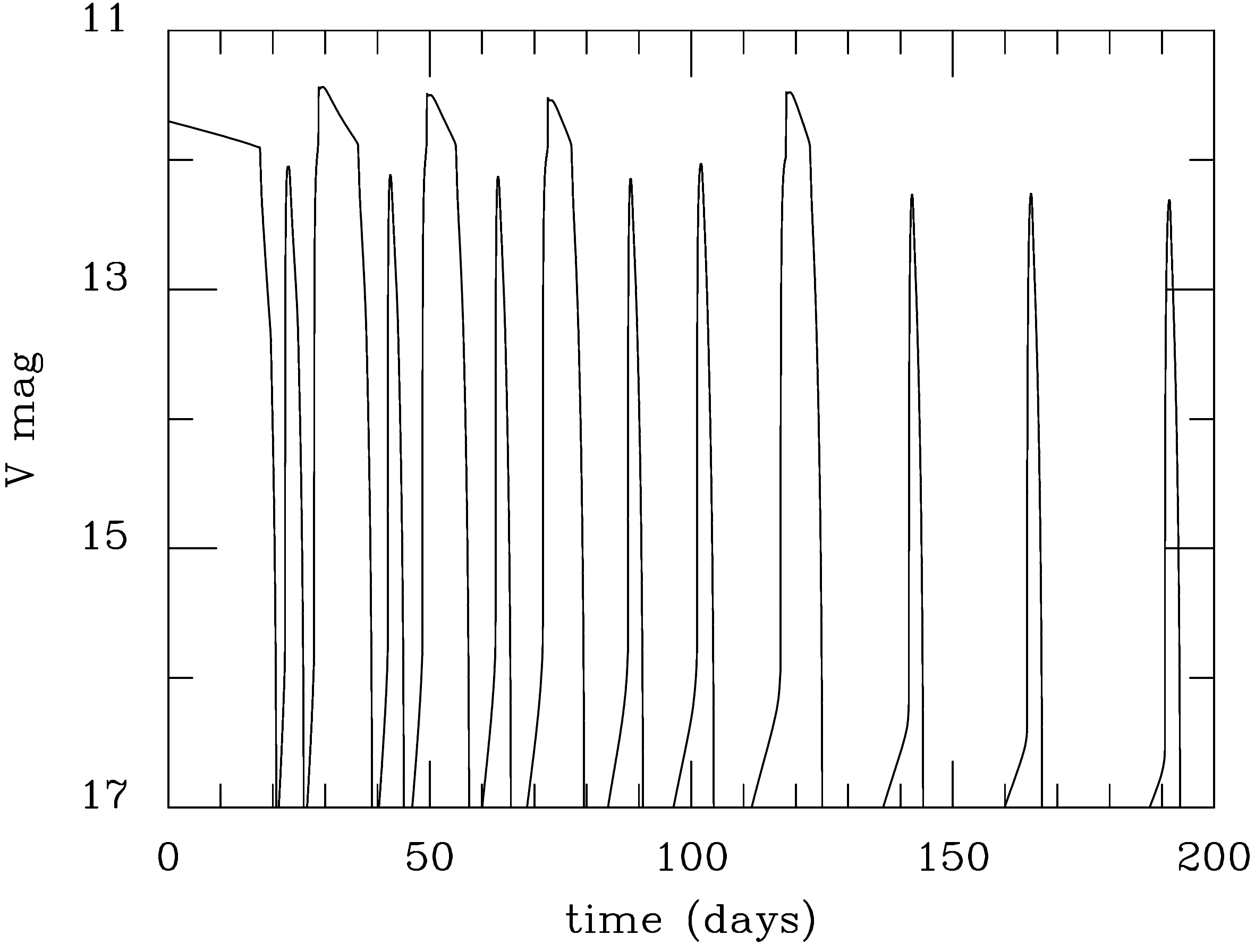}
\caption{Time evolution of a system with the same parameters as in Fig. \ref{fig:model52}, when the stream overflows the disc and 90\% of the mass is deposited at the circularization radius.} 
\label{fig:over}
\end{figure}

The first rebrightening occurs 8.8~d after the end of the main outburst, which is much longer than observed in TCP~J2104. The reason for this relatively long quiescence time is due to the fact that, although of the outside-in type, the outburst is not triggered at the very outer edge of the disc, but closer to the white dwarf so that it  takes a viscous time for matter accumulated at the outer disc, non-affected by the heating front passage, to move inwards. A shorter quiescence time can be therefore obtained if most of the transferred matter is not deposited at the outer disc (at the ``hot spot'') but if the mass-transfer stream overflows the accretion disc and deposits matter closer to the white dwarf. This is indeed what simulations show. Figure \ref{fig:over} presents the time evolution of a system with the same parameters as in Fig. \ref{fig:model52}, but in which 90\% of the transferred mass is deposited at the circularization radius, the remaining 10\% being deposited at the disc outer edge. The first quiescence interval, defined as $V > 14$ is reduced to 2.4~d. \citet{hessman99} predicted that that stream overflow should occur in quiescent dwarf novae, particularly those of the shortest orbital periods, which was confirmed by observations by \citet{smak12} of two SU UMa stars, Z Cha and OY Car. Our assumption about this effect occurring in TCP~J2104 is just fully consistent with theoretical predictions and observations. 

A 3D hydrodynamical study of the interaction of the stream with the accretion disc would be requested to determine the fraction of the mass transferred from the secondary that is incorporated in the disc at each radius; this is clearly outside the scope of this paper. We have instead followed an approach in which the number of free parameters is minimal. The ballistic stream, when it overflows the accretion disc, deposits mass at radii between the outer radius and its distance of closest approach, noted as $\varpi_{\rm min}$ by \citet{ls75}. For the low mass ratios of interest here, this distance of closest approach is of the order of 0.10 -- 0.15 times the orbital separation, about half the circularization radius. The boundary condition we use requires that a fraction of the mass transferred from the secondary is deposited at the outer edge of the disc; we use 10\% here, so that the overflow effect is dominant. As a simplifying assumption, we consider that the remaining 90\% is deposited at the circularization radius. Depositing matter at a larger radius would increase the recurrence time; however, matter deposition at a shorter distance would not change significantly the recurrence time because the first reflare is triggered at a radius that is within 10\% of the circularization radius.\footnote{Efficient irradiation probably requires a warped disc \citep{vh07} in which case the stream will reach radii smaller than the circularisation radius.}

The only other possible way to shorten the recurrence time would be to increase $\alpha_{\rm c}$, typically by a factor two to three, which would make $\alpha_{\rm c}$ uncomfortably close to $\alpha_{\rm h}$ while the DIM requires the viscosity on the hot and cold branch to differ by factors of a few for the limit cycle to produce outbursts \citep{s84}.

\begin{figure}
\includegraphics[width=\columnwidth]{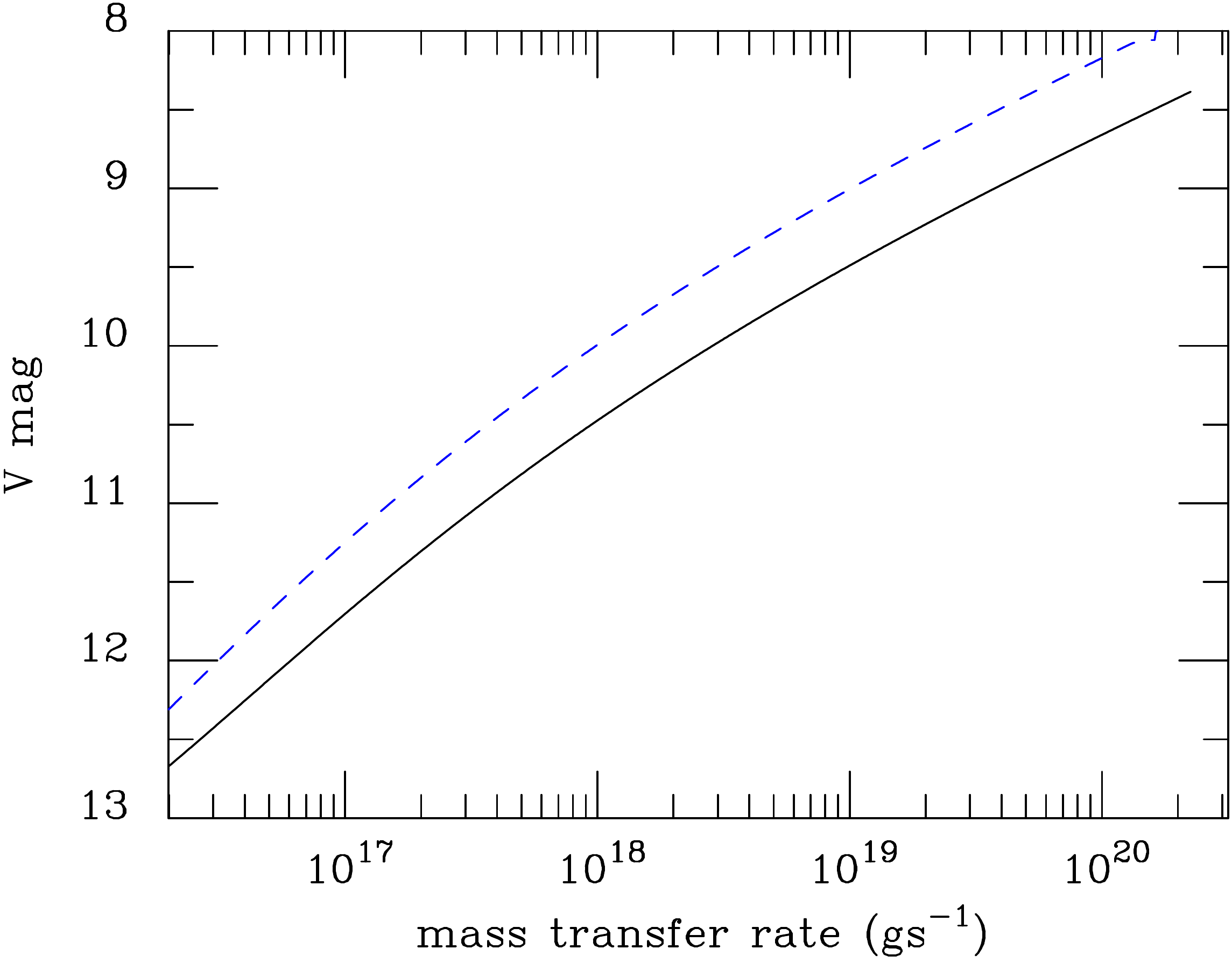}
\caption{$V$ magnitude of a steady accretion disc with the orbital parameters of TCP 2104, for $M_1=1$ (dashed blue curve) and $M_1=0.6$ (solid black curve). In both cases, $\cos i = 0.45$.}
\label{fig:steady}
\end{figure}

Irradiation of the accretion disc can also lead to rebrightenings, with two possible mechanisms. First, a cooling wave can be reflected into a heating wave; this occurs if the white dwarf is hot and the inner disc is not truncated \citep[see e.g. Figs. 5 and 6 in][]{hlw00}. In this case, the reflare starts during the decay from a main outburst, and there is no quiescence at all between the main outburst and the rebrightening, contrary to what is observed in TCP~J2104. Another possibility is that, because the critical $\Sigma_{\rm max}$ can be much reduced by irradiation, an outburst is triggered on a time scale of the order of the diffusion time in the inner disc, much shorter than the diffusion time of the full disc. In this case, the mass distribution in the outer disc has not changed much since the end of the last outburst, and a heat front cannot propagate to large distances in the disc; the outburst amplitude is thus small. This is also clearly not the case in TCP~J2104, although such a small magnitude outburst might have occurred in TCP~J2104; \citet{tni20} observed a 0.6 mag outburst that was also detected by the AAVSO and ASAS-SN on BJD 245875, 12 days after the end of the main outburst, barely visible on Fig. \ref{fig:steady}.

\subsection{Quiescence}
\label{sec:quiesc}

The nature of the quiescent phase of the TCP~J2104 outburst cycle can be deduced from the main outburst properties since, according to the standard DIM,  it is during quiescence that mass accreted during the eruption is is supposed to accumulated.

Assuming that the disc is close to being steady at the maximum of the first superoutburst, one can  determine the peak accretion rate onto the white dwarf, as well as the disc mass. Figure \ref{fig:steady} shows the disc's $V$ magnitude as a function of accretion rate of a system in steady state with the same orbital parameters as previously for two different values of the primary mass; we used $\alpha_{\rm h}=0.2$ in our simulation, but the result is independent of  viscosity. For $M_1 = 0.6$, $\dot M = 2 \times 10^{20}$~g~s$^{-1}$. Even pushing the parameters to their limits (larger disc, lower inclination), $\dot M$ still has to be higher than $5 \times 10^{19}$~g~s$^{-1}$. For a 1~M$_\odot$ primary, these numbers are reduced by a factor of approximately five. The disc mass at maximum is 10$^{25}$~g, for $M_1=0.6$ and, pushing again the parameters to their limits, larger than $10^{24}$~g. This large value of the disc mass at the onset of an outburst is reminiscent of what is found in a system such as WZ Sge, for which the disc mass, estimated at the outburst onset to be the amount of mass accumulated between outbursts is $10^{24}$~g and the peak mass accretion rate is $3 \times 10^{18}$~g~s$^{-1}$ \citep{s93}. As for WZ Sge, there are two possible explanations for this large disc mass:

(1) The viscosity is very low in quiescence, so that the disc mass can build up. Since the surface density must be less that the critical surface density $\Sigma_{\rm max}$ for the disc to stay on the stable, cold branch during quiescence, this sets an upper limit on $\alpha_{\rm c}$; using the fits given by \citet{l01} for $\Sigma_{\rm max}$, one gets
\begin{equation}
\alpha_{\rm c} < 5.0 \times 10^{-5} M_1^{0.46} \left(\frac{r_{\rm out}}{10^{10} \; \rm cm}\right)^{3.78}.
\end{equation}
For $M_1=0.6$, $\alpha_{\rm c}$ has to be less than $5.1 \times 10^{-4}$. This upper limit agrees well with the estimates provided by \citet{s93} and \citet{o95} for WZ Sge. As for WZ  Sge, one should wonder why in these systems the viscosity would so much smaller during quiescence than in systems with slightly longer orbital periods. Because of our poor understanding of the nature of angular momentum transport during low states, this argument could be circumvented, even though the binary parameters of WZ Sge and other SU UMa systems are not so different. There is here another, much stronger difficulty: the existence of three superoutbursts occurring up to almost one year after the initial outburst clearly shows that the viscosity was not that low, at least during one year. In this respect TCP~J2104 is a Rosetta stone which links  normal SU UMa stars with WZ Sge systems. One might imagine a model in which $\alpha_{\rm c}$ would slowly decrease on a very long time scale under the effect of the decreasing heating of the disc by the colder and colder white dwarf. The irradiation temperature at the outer edge of the disc for a 0.6~M$_\odot$ white dwarf is $0.07 T_{\rm wd}$, too small to have a significant influence on, for example, the ionisation properties of the accretion disc at a time when the last observed superoutburst occurs and $T_{\rm wd}$ is of the order of 10$^4$K, making such a model rather useless.

(2) $\alpha_{\rm c}$ in quiescence is of the order of $\alpha_{\rm h}/10$, as in normal CVs. The inner disc is truncated to a radius large enough that the disc can be (quasi--)steady on the stable, cold branch with a relatively low mass transfer rate, typically of the order of few times 10$^{15}$~g~s$^{-1}$; the outburst has then to be triggered by an enhancement of mass transfer from the secondary. This explanation was put forward by \citet{hlh97} for WZ Sge. This explanation is also strengthened by the fact that models in which the quiescent disc is unsteady often predict disc magnitudes brighter than the observed $V\sim 18$; typically $V\sim 16 - 17$ (see for example Fig. \ref{fig:model52}) or even brighter if the disc is truncated. One should, however, keep in mind that our understanding of low states in CVs, and in particular of disc spectra at low accretion rates is rather limited, and this argument in favour of a stable, steady quiescent cold discs is by far not compelling.

This being said, we adopt the model of \citet{hlh97} in which, during deep quiescence, the disc is stable on the cold branch. This requires the mass transfer rate to be less than the critical value at the \textsl{inner edge} of the disc, and that the disc is truncated by, for example, the magnetic field of the white dwarf. Truncation of the inner disc can also occur if there is a transition from the geometrically thin, optically thick flow to an optically thin, geometrically thick flow as in  the case of an ADAF (advection dominated accretion flow) or because of mass ejection \citep[see e.g. models by][]{mm94,lmm97}. In any case, discs are observed to be truncated in quiescent DNe \citep{b15}, with radii of the order of $(3 - 10) \times 10^9$~cm. 

Here, we focus of the simple case of truncation by a magnetic field at a radius given by the Alfvén radius:
\begin{equation}
r_{\text {in }}=5.1 \times 10^{8} \mu_{30}^{4/7}\left(\frac{\dot{M}_{\mathrm{acc}}}{10^{16} \mathrm{~g} \mathrm{~s}^{-1}}\right)^{2/7} M_1^{-1/7} \mathrm{~cm},
\label{eq:trunc}
\end{equation}
where $\mu_{30}$ is the magnetic moment in units of $10^{30}$~G~cm$^3$. This magnetic truncation should be considered as a proxy for any mechanism removing the inner disc regions. As we shall see in the next section, the available data for TCP~J2104 do not constrain the mechanism that truncates the disc; this is not the case for other systems discussed in this paper. The mass transfer rate during quiescence is then limited to
\begin{equation}
\dot M_{\rm tr} < 2.5 \times 10^{14} M_1^{-0.72} \mu_{30}^{0.86} \; \rm g\, s^{-1},
\end{equation}
where we have used the fits from \citet{l01} for the critical accretion rate, and we have introduced a multiplicative factor of about four, determined by our numerical simulations, in order to account for stabilizing effect of the boundary condition $\Sigma = 0$ at the inner radius, $\Sigma$ being the surface density.

\begin{figure}
\includegraphics[width=\columnwidth]{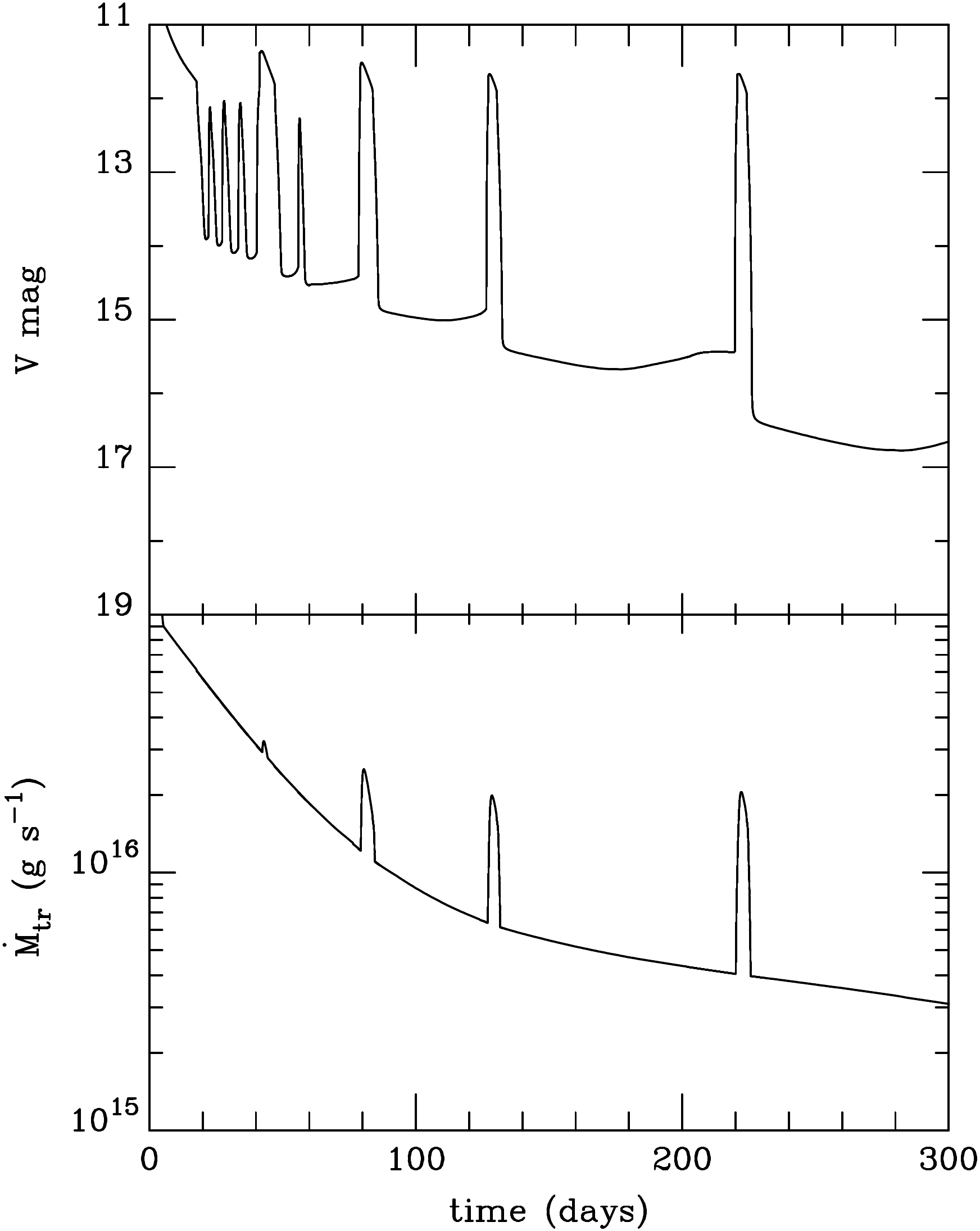}
\caption{Time evolution of a system which resembles to TCP~J2104 (see text). {\it Top panel:} $V$ magnitude, including contributions from the white dwarf and the hot spot. {\it Bottom panel:} mass transfer rate from the secondary.}
\label{fig:full}
\end{figure}

\subsection{Modelling the full light curve}
\label{sec:fullc}

Using the models of the particular phases of the TCP~J2104 outburst presented in the previous sections one can now build a model of its full outburst-cycle light curve.
Figure \ref{fig:full} shows the results of a model in which the disc is initially steady with a mass accretion rate of $2 \times 10^{18}$~g~s$^{-1}$ with $\dot{M}_{\rm tr}$ slowly decaying as shown in the bottom panel of Fig. \ref{fig:full}; the spikes in $\dot{M}_{\rm tr}$ are due to irradiation--induced enhancements of mass transfer that we describe with $\gamma = 0.2$. The stream initially overflows the accretion disc, and, as previously, we assume that 90\% of the transferred mass is incorporated at the circularization radius; when the mass transfer drops below $2 \times 10^{16}$~g~s$^{-1}$, all of the mass is added at the disc outer edge. The orbital parameters are the same as in Section \ref{sec:initial}, and we assume that the white dwarf temperature remains constant during the initial outburst at $T_{\rm wd} = 3.5 \times 10^4$~K, and decay exponentially at t $t=20$~d, when the start of a cooling front ends the initial superoutburst, with a characteristic time scale of 200~d. The disc is truncated by a magnetic field (or equivalent) with a magnetic moment $\mu = 5 \times 10^{30}$~G~cm$^3$. Truncation is required to avoid small outbursts (about 1 mag), and also to enable the system to be stable on the cold branch after 220~d. The truncation radius during quiescence is of the order of $3 \times 10^9$~cm during quiescence, comparable to what has been observed in other systems. We include contributions from the accretion disc, the white dwarf, assuming a blackbody spectrum, and from the hot spot, assuming a blackbody spectrum with a temperature of $10^4$~K, the hot spot luminosity being determined as in \citet{hkl20}. For simplicity, we add this contribution even when the stream overflows the disc; in this case, the interaction between the stream and the disc still results in additional optical light, but the emitted spectrum is certainly different from that of the hot spot. We find, however, that the hot spot contribution in the $V$ band is always less than the white dwarf contribution, so that the error introduced by this assumption is small.

The light curve we obtain, although not identical, is very similar to the observed light curve. In particular, we obtain a sequence of normal and superoutbursts, with a very short quiescence time at the beginning of the sequence, an increasing quiescence period between outbursts, the system being stable after 220 days. Because there are many parameters that are not well constrained, such as the time-variation of the mass transfer rate or the way matter is added to the disc, it does not make not sense to try to reproduce the details of the observed light curve. 

We emphasize again that in this model, the initial trigger of the outburst is not directly a disc instability but is linked to the secondary star. The mass transfer rate must, for some reason, sharply increase to explain the initial outburst. Various models have been proposed for such mass transfer outbursts. For example, \citet{warner88} argued that the magnetic activity of the secondary might account for variations of the mass-transfer rate in  cataclysmic variables; \citet{b69,b75} proposed that a dynamical instability arising in the envelope of the secondary could be responsible for the outbursts of dwarf novae, but we now know that such an instability does not exist \citep[see e.g. the discussion in][]{s84a}. The mass-transfer instability model is no longer considered as a viable model for DNe, but was suggested by \citet{hrt98} as a possible explanation for faint, stunted outbursts occurring in old novae and nova-like systems and by \citet{hl17} for faint outbursts occurring in intermediate polars. Irradiation of the secondary, not included in the above-mentioned mass-transfer outburst models, most probably plays an important role, especially because the secondary surface temperature is quite low in WZ Sge systems. However, this is not just the mass-transfer instability model revived, because the enhanced mass-transfer brings the disc to an unstable state whose evolution is the source of the observed outbursts.

\section{Application to other systems}

Our model for the observed reflares in TCP~J2104 can  be applied to other members of the WZ Sge subclass, as well as to soft X-ray transients that have several similarities with the WZ Sge stars \citep{k04}. We consider here two emblematic systems: WZ Sge, that is the prototype of its subclass, and EG Cnc, the so-called ``king of the echo outbursts'' \citep{pks98} that exhibited spectacular reflares. Our objective here is to show that a broad variety of rebrightenings can be reproduced, provided we add to the model the few ingredients we discussed above. 

\begin{figure}
\includegraphics[width=\columnwidth]{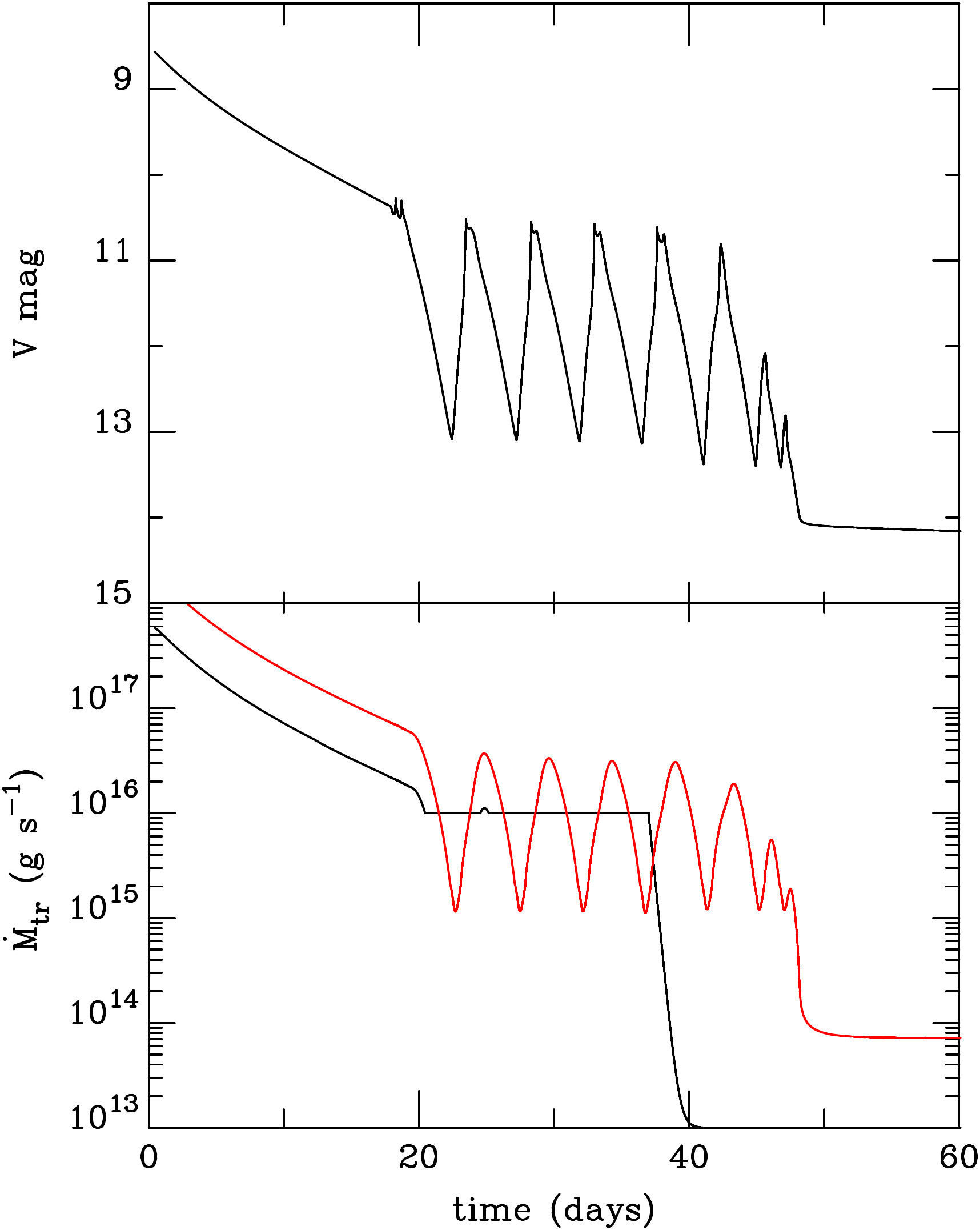}
\caption{Same as Fig. \ref{fig:full} for the parameters of WZ Sge. The red curve in the bottom panel shows the accretion rate onto the white dwarf.}
\label{fig:wz}
\end{figure}

\subsection{WZ Sge}

WZ Sge has an orbital period of 1.36~h; its distance, as determined by GAIA is 45.1~pc with an uncertainty smaller than 1~pc \citep{g18}; the inclination is well constrained by the existence of eclipses, and is close to 75$^\circ$ \citep{s93}.

WZ Sge has had four historical outbursts observed in 1913, 1946, 1978, and 2001. The last three outbursts showed a sharp decline (dip) in luminosity after about 20 to 40 days followed by a sequence of $\sim$ 1 mag rebrightenings (echoes) with a recurrence time of two days \citep{pmr02}; there are not enough data for the 1913 outburst to see it it had similar behaviour. The hot spot during the 2001 outburst was brighter by a factor of about sixty than the quiescent value, implying that enhanced mass transfer plays a major role during the eruption \citep{pmr02}; Doppler maps also indicated the importance of mass transfer enhancement \citep{s04}.

UV observations show a cooling white dwarf, from an initial temperature of 28,000~K to 16,000~K, 17 months after the outburst \citep{gsc04,lsg04}. Modelling the white dwarf cooling that remains hotter than its deep quiescence value by about 1500~K three years after the end of the outburst requires a massive white dwarf (typically 0.9~M$_\odot$), as well as continuous accretion during quiescence at a relatively high rate, of the order of $1-2 \times 10^{15}$~g~s$^{-1}$ \citep{gsc06}. Also \citet{shk07} deduce from radial velocity measurements and estimates of the gravitational field of the white dwarf that it must be massive. The primary mass required is twice the value determined by \citet{s93} and subsequently used by \citet{hlh97}, but the same value of the quiescent mass-transfer rate was deduced from optical observations by \citet{s93}. There is little doubt that WZ~Sge accretes in quiescence since it emits X-ray with a luminosity $(1\, - \, 2) \times 10^{30}$~erg~s$^{-1}$ \citep{nk14}. 

Mass transfer rates as high as $10^{15}$~g~s$^{-1}$ exceed by far the critical rate for a disc reaching the white-dwarf surface to be stable on the cold branch. For such an accretion rate the disc, to be in quiescence, must be truncated at a radius of a few $10^9$~cm, much larger than the white dwarf radius. \citet{hlh97} found that with a truncation radius given by the Alfvén radius with $M_1 = 0.45$ and $\mu_{30}= 10$, the quiescent disc of WZ Sge is cold, steady (and stable). In such a truncated disc, the accretion and mass-transfer rates are equal and correspond to the observed X-ray luminosity. This model of WZ Sge quiescence \citep{hlh97} is the basis of the option (2) chosen for TCP~J2104 in Section \ref{sec:quiesc} and requires the WZ Sge superoutburst to begin with a rapid enhancement of mass-transfer from the secondary star. One expects therefore that the model of Section \ref{sec:fullc} should also apply to  the description of the light curve of the subclass prototype. A quick look at the observations of these two systems, however, suffices to show that this application will not be straightforward since the light curves of the two systems are different in many important aspects. The main two differences are: i.) there is only one (super)outburst in WZ Sge, versus six in TCP~J2104; ii.) the twelve rebrightenings \citep[called ``echoes'' in][]{pmr02} around a ``standstill'' in WZ Sge have no equivalent in the light curve of the other system. We first try to find what ingredient is needed to produce the echoes.

The striking characteristic of the ``echoes'' in WZ Sge is that there is no quiescent period: a rebrightening is initiated before the previous one has ended, implying that the cooling front that propagates inwards is reflected into a heating front that brings the disc fully (or almost fully) into a hot state. Such reflections require, or are favoured by, several ingredients: a massive compact accreting object (small radius), a hot white dwarf, and a small inner disc radius; they are in particular a natural outcome of models for SXTs with no disc truncation \citep{dhl01}.

We model a system with the same orbital period as WZ Sge,  masses $M_1 = 1$,  $M_2=0.07$ and $i=75^\circ$. For reasons explained below, the white dwarf magnetic moment is taken to be $0.5 \times 10^{30}$~G~cm$^3$; we use $\alpha_h=0.2$ and $\alpha_c=0.04$. We assume that the white dwarf temperature exponentially decreases from a peak-of-outburst temperature of 28,000~K with a decay time of 400~d. The parameter $\gamma$ is set to 0.3. As for TCP~J2104, the disc is almost steady during the outburst, with a high mass transfer rate from the secondary, but here, if we wish to reproduce echoes, we need the mass--transfer  to stay for about twenty days roughly constant at some intermediate rate, before dropping to a low value. As can be seen in Fig. \ref{fig:wz}, the light curve we obtain is similar to that of WZ Sge during its 2001 outburst, although the oscillation period of the reflares is longer than observed, being of the order of four days instead of the observed two days. The oscillations are caused by the the reflection of the cooling front into a heating front close to the surface of the white dwarf (that is why we need a feeble magnetic moment), as a result of irradiation by the hot white dwarf. Their oscillation period therefore depends mainly on the disc size, which is tightly constrained, and on the propagation speed of the cooling front, determined by $\alpha_h$.  Reducing this oscillation period can be achieved by increasing $\alpha_{\rm h}$ by a factor two. Here, different from TCP~J2104, the rebrightening period is not influenced by the way matter is incorporated into the disc since the underlying mechanism is different, consistent with the fact that the light-curve form is also different. As mentioned in Section \ref{sec:rbr} the front reflection produces light curves with no quiescence, as observed in WZ Sge, but not in TCP~J2104.

The model producing the light curve presented in Fig. \ref{fig:wz} also predicts two normal outbursts, not shown on this figure, occurring at times $t=135$ and $t=280$~d after the start of the superoutburst, before the system returns to full quiescence. These outbursts are due to the fact that, although the mass transfer rate is low, the disc mass is high enough that viscous diffusion can bring the surface density above the critical value close to the inner edge of the disc. The disc is close to being stable: the mass lost during each of these two outbursts is of the order of 10\% of the total mass, and reducing the disc mass by 20\% is sufficient to render the subsequent evolution stable. Such outbursts have not been detected but the reason for their appearance, the low magnetic moment allowing for small inner disc radii in quiescence is also incompatible with the required large quiescent disc truncation, as discussed above (see Eq. \ref{eq:trunc}).

It is possible that a better tuning of the parameters ($M_1$, $\mu_{30}$, $T_{wd}$) makes it possible to avoid secondary outbursts, but a more promising possibility, is that, for low mass accretion rates, typically less than $10^{15}$~g~s$^{-1}$, the truncation radius is larger than the magnetospheric radius, meaning that disc truncation is not magnetic but due to evaporation \citep[see also][]{mhml98}. This would enable the system to be stable on the cold branch at the end of the reflare period without perturbing the reflare sequence.

This modification of the truncation radius would also produce the accretion rates required to account for the slow cooling of the white dwarf being at the same time compatible with the observed quiescent X-ray luminosity. In other words, what we propose here is an ``improved'' version of the \citet{hlh97} model in which we account for both the main outburst and the reflares. The improvement is put in quotation marks because the pause in the mass-transfer decay is still left to be explained.

\begin{figure}
\includegraphics[width=\columnwidth]{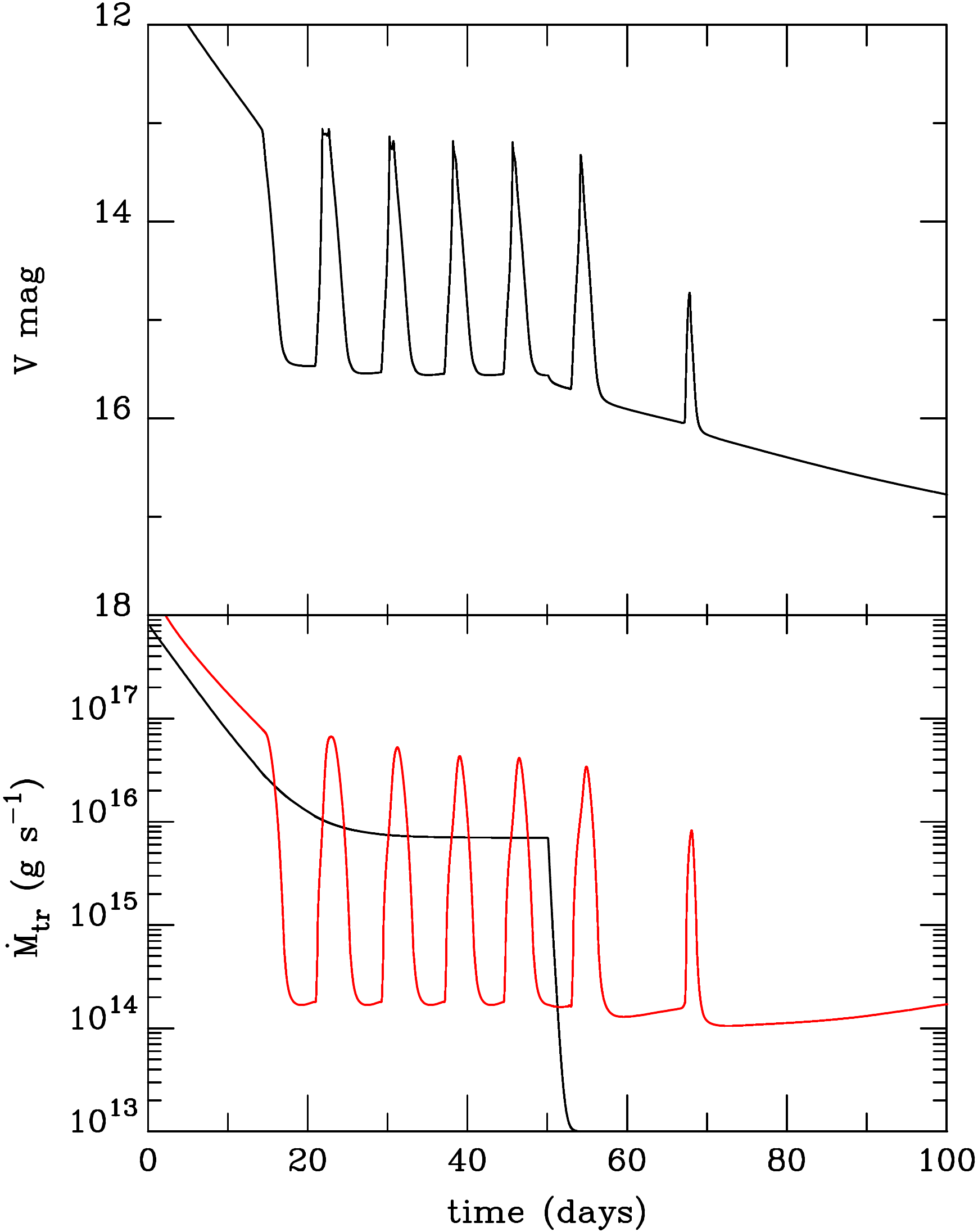}
\caption{Same as Fig. \ref{fig:wz} for the parameters of EG Cnc.}
\label{fig:eg}
\end{figure}

\subsection{EG Cnc}

EG Cnc was discovered on archival plates during an outburst that occurred in 1977 \citep{h83}, and since then it had two superoutbursts, one in 1996-1997 \citep{pks98} and one in 2018 \citep{kikk21}; it underwent what appears to be a faint, normal outburst in 2009 \citep{t09}. The light-curves of the 1996-1997 and 2018 outbursts are almost identical \citep{kikk21}; both show six rebrightenings that resemble normal outbursts with an average recurrence time of one week. They are reminiscent of the first reflares of TCP~J2104, with the difference that they consist only of normal outbursts, and that the recurrence time between outbursts is almost constant. This sequence resembles a normal sequence of outbursts in an almost relaxed situation; it should thus be possible to reproduce such a sequence with a constant mass transfer rate which drops rapidly at the end of the reflare period. As for TCP~J2104, the quiescent luminosity decreases with time, which could be accounted for by the white dwarf cooling. We show in this section that, with these ingredients and slightly  different parameters from TCP~J2104, one can produce light curves that are very similar to that of EG~Cnc.

The orbital system of EG Cnc is 1.439~h; its distance, as determined by GAIA EDR3, is 187~pc, with an uncertainty of the order of 7~pc \citep{g18}, significantly smaller than the distance of 320~pc determined by \cite{pks98}. The mass ratio, as determined by the superhump period, is in the range 0.027 -- 0.035 \citep{pks98,kikk21}; the primary mass is essentially undetermined. We take here $M_1 = 0.6$~M$_\odot$, and $M_2=0.06$~M$_\odot$, higher than the mass inferred from the superhump period (0.016 -- 0.021~M$_\odot$) which is very low, even for post-period minimum systems \citep[see e.g. models by][]{kb99}. In any case, models are weakly dependent on $M_2$. Apart from the absence of eclipses, the inclination is not constrained; we take here $i=63^\circ$ which reproduces well the magnitude at which the initial outburst abruptly declines. We use $\alpha_{\rm c} = 0.2$ and $\alpha_{\rm c} = 0.04$, as in previous simulations.

We start at time $t=0$ with a steady accretion disc with a mass transfer rate of $2 \times 10^{18}$~g~s$^{-1}$. At this moment, we set abruptly the mass transfer rate equal to $8 \times 10^{17} \exp (-t/4 \; \rm d) + 10^{16}$~g~s$^{-1}$, until $t=50$~d, at which point it drops to $10^{13}$~g~s$^{-1}$. This mimics the evolution of $\dot M_{\rm tr}$ during the initial outburst, presumably driven by a short, impulsive mass transfer outburst maintained for a while by the irradiation of the secondary which is accounted for by the explicit exponential time dependence instead of the $\gamma$ term, and by the subsequent time evolution with a plateau during which the reflares should occur. We assume an initial white dwarf temperature of $2.8 \times 10^4$~K that remains constant for 50~d and then decreases exponentially with an e-folding time of 100~d.

Figure \ref{fig:eg} shows the light curve we obtain with these parameters. The magnetic moment is $5\times 10^{29}$~G~cm$^3$. We do find a sequence of reflares; the corresponding outbursts are  intermediate between outside-in and inside-out, being triggered at a distance from the white dwarf that is $2.5 \times 10^9$~cm, twice the inner disc radius. This, together with the irradiation of the inner disc by the hot white dwarf accounts for the short recurrence time, of about eight days, as observed. A relatively small truncation radius is required for irradiation to be efficient, hence the small value of the magnetic moment. During the period of reflares, the disc is not far from being relaxed; about 15\% of its mass is accreted during an outburst, and the number of reflares essentially depends on the duration of the plateau phase for $\dot M_{\rm tr}$ since the white dwarf temperature varies on a comparatively longer time scale.

As for the WZ Sge model, there is an additional outburst that occurs at time $t=112$~d, not shown on the figure, and that has not been observed. Although it is possible that such an outburst could have been missed, since the AAVSO data show that the coverage of this source was quite sparse at that time, it is probably safer to consider that this outburst has not happened. Simply assuming that the disc inner radius does not follow the dependence on accretion rate given by the Alfvén radius, as was done in the case of WZ Sge, will not suffice to solve the problem because the accretion rate between two reflares is very similar to the accretion rate during the full quiescence. However, evaporation of the inner disc, as, for example, suggested by \citet{mm94} would not only increase the inner disc radius but a time-dependent process would also decrease the surface density, which could be a solution to this problem. As for WZ Sge, it would also avoid the need for a very low quiescent mass transfer rate.

\subsection{Soft X-ray transients}

Reflares, well separated from the main outburst, are also sometimes observed in SXTs. GRO J0422+32 is one of the best documented cases; its 1992 outburst was followed by two echo outbursts occurring about 150 and 250 days after the termination of the main outburst \citep{cgm95,ci95}, which lasted for 20 to 30 days, much less than the duration of the main outburst, of the order of 200 days.  \citet{khv96} suggested that these echo outbursts could be due to the same mechanism that produces echoes in WZ Sge stars, the longer time scales being being explained by a much larger discs in SXTs, and hence much longer viscous time. However, the thermal-viscous instability in SXTs is strongly modified by X-ray irradiation of the accretion disc, and the details of irradiation are not well understood: reprocessing of the X-rays on the outer disc is observed to dominate the optical light, and yet simple models predict that self-screening should prevent  X-rays to reach the disc outer parts \citep[see e.g.][]{dlh99}. X-ray irradiation is therefore modelled in a phenomenological way, and adds at least one additional free parameter to the model \citep{tdl18}. In these circumstances, reproducing the observed echo outburst is unlikely to bring much insight on the physics involved in SXTs.

During SXT outburst decay, one sometimes observes so called glitches:  the X-ray luminosity suddenly increases by a factor of typically two to three. These glitches are a natural outcome of the DIM when irradiation is important and the disc is not, or not enough, truncated \citep{dhl01}. It has also been suggested that glitches are caused by an increase of the mass transfer rate as a result of the secondary being irradiated by X-rays \citep{aks93}, but, although mass transfer does increase during outbursts, disc models have not reproduced the sharp glitches that are observed: the increase of the mass transfer rate is smoothed on the long viscous time of the disc. Reflares and glitches in SXTs show that the standard irradiated DIM cannot fully reproduce the observed light curves of these systems, but also that trying to remedy this imperfection by ad hoc added ingredients is rather pointless.

\section{Conclusions}

We have shown that the very unusual light curve of TCP~J2104 can be explained by the DIM provided that its standard version is enriched by the inclusion of several ingredients related to time-variations of the parameters entering the DIM. The mass transfer rate from the secondary is increased during the main outburst to values comparable to the mass accretion rate and then exponentially decreases with time, remaining high for a few months before returning to its low quiescence value. We also require the stream from the secondary to overflow the accretion disc when the mass transfer rate is high; this reduces the diffusion time of the disc region involved in outburst, and accounts for the short recurrence time of the echo outbursts appearing shortly after the end of the main outburst. During deep quiescence, the truncated disc is stable on the cold branch; The truncation is either magnetic or due to some other mechanism; in the case of TCP~J2104, there is no strong constraint on the truncation physics. The long-term decay of the quiescent luminosity is accounted for by the cooling of the white dwarf, which is the main contributor to the optical light during quiescence, and, to a lesser extent, to the decrease of the mass transfer rate and hence of the hot spot luminosity. The predicted magnitudes, well constrained thanks to an accurate determination of the system distance by {\it GAIA} are in good agreement with observations.

Echo outbursts that are often detected in other WZ Sge stars can also be explained by the DIM using the similar ingredients. We have been able to reproduce light curves resembling those of WZ Sge and EG Cnc; this requires that the truncation radius of the inner disc is not the Alfvén radius, but increases more rapidly with decreasing mass accretion rate in the case of WZ Sge, or, in the case of EG Cnc, that the mechanism that truncates the disc also affects the surface density in the inner parts the disc. This is needed in order to avoid one or two additional reflares that would otherwise occur a few months after the system has entered quiescence, which is not observed. This argues in favour of disc truncation by evaporation of the inner disc, which leads either to the formation of a wind or to the formation of an optically thin accretion flow similar to the ADAFs discussed in the context of low-mass X-ray binaries.

Echo outbursts such as observed in GRO J0422+32 after a quiescent period of a few months could in principle be explained using the same ingredients. Although it should be easy to produce light curves with echo outbursts using a decreasing mass transfer rate after an initial increase during the outburst peak, we refrain from doing so because the DIM in SXTs is strongly modified by X-ray irradiation of the disc, which introduces an additional parameter that is poorly constrained. We suggest that glitches that occur in several SXTs during decay would rather be due to a reflection of the cooling wave into a heating wave that is naturally found in the DIM when irradiation of the disc is included and the inner disc is not truncated to a large value.

Our model relies on several important hypotheses. First, that the value of viscosity parameter $\alpha$ in quiescence in WZ Sge systems is the same as in other dwarf novae, as suggested by \citet{hlh97}. The existence of normal and superoutbursts in TCP~J2104 clearly support this conservative assumption, at least during one year after the initial superoutburst. From this, it follows immediately that WZ Sge systems must be stable on the cold branch during quiescence and that their large outbursts (superoutburst) must be triggered by a mass transfer enhancement event from the secondary. The physical cause of this mass transfer burst remains to be found. It also follows that the disc must be  severely truncated in quiescence, either  by the presence of a magnetic field or by the formation of a hot, optically thin flow close to the white dwarf. There are observational evidences of this \citep[see e.g.][]{b15}; we have shown that, whereas the rebrightenings in TCP~J2104 can be explained by both truncation mechanisms, other systems require that the truncation radius not to be the Alfvén radius and an evaporation mechanism is thus preferred. The second, important hypothesis is that mass transfer must be sustained, possibly during and certainly at the end of the initial large outburst; this can be a result of the (unspecified) mechanism that triggers the mass transfer outburst; also, irradiation of the secondary star by the hot white dwarf is certainly an important ingredient. Finally, we requite that the stream overflows the accretion disc at times when the mass transfer rate is high, resulting in mass being deposited at distances closer to the white dwarf than the disc outer radius.

With these three hypotheses, we have been able, for the first time, to explain the rebrightenings observed in WZ Sge systems. For doing so, we have made, by necessity and in order to limit the number of free parameters, a number of assumptions on how these hypotheses are implemented. The detailed parametrization is an oversimplification of the complex physics at work in these systems which prevents us to draw more specific conclusions on, for example, the exact value of our parameters. The very fact that, using plausible parameters, we have been able to reproduce the general shape of the observed light curves as well as several characteristic observed values such as magnitudes, outburst duration, time interval between rebrightenings, etc. is nevertheless a strong support to the model.

The model we have used here, although successful in reproducing the observed light curves, suffers from several drawbacks. We have assumed that the disc has a axial symmetry, which is incorrect in its outer parts, in particular for systems with low mass ratios such as the ones considered here. Even if the 3:1 resonance does not induce a tidal instability as proposed by \citet{o89}, the propagation of heating or cooling fronts in the outer disc is certainly perturbed by the disc eccentricity and non--planarity. Moreover, disc warping has a major influence on the irradiation efficiency as well as on the interaction of the stream from the secondary with the disc itself. One must also emphasize that the disc structure during quiescence, in particular in the WZ Sge stars in which the mass transfer rate and consequently the disc temperature are low, is not well established. There are many uncertainties on the basic physics of discs in quiescence, in particular with regards to the determination of opacities at low temperatures, and on the nature of angular momentum transport when the disc is so cold that is if fully neutral \citep{gm98,sldf18}. Finally, winds have not been included in the model, except by assuming that the inner disc is truncated by some mechanism which could be the formation of a wind. Winds have other important effects: they carry mass and consequently angular momentum which can affect the shape of the light curve in particular in the case of SXTs in which they are strong \citep{tlh18,ddt19}, but they can also, if they are magnetized, be the main source of angular momentum transport in the inner disc \citep{sdl19,sld20}. Finally, despite serious progress in understanding the physics of CV discs \citep{ckblh16,jsz16,jsz17,sldf18,sdl19}, to describe dwarf--nova outbursts we are still forced to use the $\alpha$--prescription disc model, with all its limitation.

Our models of rebrightenings, reflares, and echoes in dwarf nova outbursts should therefore be considered as phenomenological illustration of the physical ingredients that must be included into models of the disc thermal-viscous-instability driven outburst. 

\begin{acknowledgements}
We thank Joe Smak for moral support. We acknowledge with thanks the variable star observations from the AAVSO International Database contributed by observers worldwide and used in this research. JPL was supported 
in part by a grant from the French Space Agency CNES.
\end{acknowledgements}

\bibliographystyle{aa}
\bibliography{biblio}

\end{document}